\documentclass[preprint,aps,amssymb,superscriptaddress,nofootinbib]{revtex4}
\usepackage{graphicx}
\usepackage{braket}
\newcommand{\del}{\partial}
\newcommand{\m}{\mathring}

\begin{document}

\title{Quantum geometry with a nondegenerate vacuum: a toy model}

\author{Sandipan Sengupta}
\email{sandipan@rri.res.in}
\affiliation{Raman Research Institute\\
Bangalore-560080, INDIA.}

\begin{abstract}
 Motivated by a recent proposal (by Koslowski-Sahlmann) of a kinematical representation in Loop Quantum Gravity (LQG) with a nondegenerate vacuum metric, we construct a polymer quantization of the parametrised massless scalar field theory on a Minkowskian cylinder. 
 The diffeomorphism covariant kinematics is based on states which carry a continuous label corresponding to smooth embedding geometries, in addition to the discrete embedding and matter labels. 
 The physical state space, obtained through group averaging procedure, is nonseparable. A physical state in this theory can be interpreted as a quantum spacetime, which is composed of discrete strips and supercedes the classical continuum. We find that the conformal group is broken in the quantum theory, and consists of all Poincare translations. These features are remarkably different compared to the case without a smooth embedding. Finally, we analyse the length operator whose spectrum is shown to be a sum of contributions from the continuous and discrete embedding geometries, being in perfect analogy with the spectra of geometrical operators in LQG with a nondegenerate vacuum geometry.
\end{abstract}


\maketitle

\section{Introduction}
In relativistic field theories, the action integral is usually given in flat spacetime, where the dynamics of the fields is parametrised by inertial coordinates. There exists a different description of the same theory in terms of arbitrary spacetime coordinates. Within this formulation, the inertial coordinates are treated as dynamical fields, resulting in an enlarged set of phase space variables whose motions are parametrised by arbitrary coordinates. Since the dynamical content of the theory does not depend on the choice of these coordinates, this leads to a generally covariant formalism. 

The original idea of such a reformulation (`parametrisation') of field theories was due to Dirac\cite{dirac}, and was used by Kuchar\cite{kuchar} extensively in his analysis of free scalar field theory in a two dimensional flat spacetime of cylindrical topology, also known as Parametrised field theory (PFT). Subsequently, he constructed a Dirac quantization of this model, and used it as a toy system for canonical quantum gravity\cite{kuchar1,varadarajan}. More recently, through Laddha and Varadarajan's analysis of PFT within the Loop (polymer) quantization framework\cite{l,laddha,laddha1}, this system has emerged as a robust testing ground for the quantization techniques employed in loop quantum gravity. The remarkable fact that two dimensional PFT can be quantized and solved using the LQG-based quantization approach suggests that the insights gained in this context might be useful in the analysis of four-dimensional gravity theory, for which such a success still remains elusive. In fact, the Hamiltonian constraint in PFT was studied precisely from such a perspective\cite{laddha2}.

The action for free scalar field theory in $1+1$ dimensional Minkowskian spacetime is given by:
\begin{eqnarray}\label{s}
S[\phi(X)]~=~\int d^2 X ~\eta^{AB}\del_A \phi(X)\del_B \phi(X)
\end{eqnarray}
where, $\phi(X^A)$ are the (dynamical) scalar fields; $X^A,~A=0,1,$ are the (non-dynamical) inertial coordinates corresponding to the flat spacetime; $\eta_{AB}\equiv [-1,1]$ is the flat metric. This system can be `parametrised' by introducing a set of curvilinear coordinates $x^\alpha=(t,x)$ and treating the $X^A(x)$-s as dynamical fields alongwith $\phi(x)$, such that for fixed $t$, the variables $X^A(t,x)$ define a spacelike surface. The motions of $X^A(x),\phi(x)$ are now parametrized by $x^\alpha$, and is determined by the following action principle:
\begin{eqnarray}\label{s-pft} 
S_{PFT}[X^A(x),\phi(x)]~=~\int d^2 x ~g(X)^{\frac{1}{2}}g^{\alpha\beta}(X)\del_\alpha \phi(x)\del_\beta \phi(x)
\end{eqnarray}
Here $g(X)$ is the determinant of the metric $g_{\alpha\beta}(X)$, given by:
\begin{eqnarray*}
g_{\alpha\beta}(X)~=~\eta_{AB} \frac{\del X^A(x)}{\del x^\alpha}\frac{\del X^B(x)}{\del x^\beta}
\end{eqnarray*}
The action (\ref{s-pft}) exhibits general covariance. Although (\ref{s-pft}) corresponds to an enlarged phase-space, the two-functions worth of gauge freedom in the choice of coordinates $x^\alpha$ implies the presence of a pair of first-class constraints in the canonical theory, thus leading to the same number of physical degrees of freedom as in the original theory given by  (\ref{s}) (for details, see \cite{kuchar}).

In the polymer quantization of this theory\cite{laddha,laddha1}, the basic operators are chosen to be the analogues of the holonomy and flux operators as used in canonical LQG kinematics. In such a formulation, the elementary excitations are one-dimensional polymer-like objects, characterised by coloured graphs, whose edges carry integer-labelled charges\cite{laddha1}. These are known as charge-network states, and define the kinematical states in the theory. The (kinematical) `vacuum' state is given by the empty state without any charge-network. Thus, the vacuum geometry is characterised by a degenerate metric. Again, these states are very similar to the kinematical states used in LQG, known as the spin-network states. There also, the `vacuum' is just the empty state, corresponding to zero eigenvalues of the area and volume operators. The full spectra of these geometrical operators provide an elegant description of the underlying discreteness of spatial geometry\cite{ashtekar,rovelli}. 
However, within such a framework, there is no natural way to deal with asymptotically flat spacetimes. In other words, using the LQG kinematics simply consisting of one dimensional excitations on top of a degenerate vacuum, it is difficult to implement the boundary conditions at asymptotia where one expects the spacetime geometry to become smooth. One way to get around this problem would be to modify (minimally) the kinematical representation as used in LQG.

Recently, just such a representation for LQG was proposed by Koslowski \cite{k} and further studied by Sahlmann\cite{s}, where the kinematical construction involves nondegenerate spatial background metrics. This was motivated from a view to set up a diffeomorphism invariant framework which can admit an effective description of smooth spacetimes. 
In this formalism, each state carries a continuous label corresponding to smooth spatial geometries\footnote{The background, and hence the corresponding continuous label of the states, is not fixed and changes under the action of the constraint operators, in particular under diffeomorphisms. This leads to a diffeomorphism-covariant quantum kinematics, and is consistent with the spirit of a background-independent quantization. }, in addition to the usual spin-network labels. The action of the flux operators on these states is different as compared to its action in the standard (Ashtekar-Lewandowski) representation, and has an additional piece coming from a classical background geometry\cite{s}:
\begin{eqnarray*}
\hat{E}_{s,f}\Psi~=~X_{s,f}\Psi~+~\m{E}_{s,f}I \Psi
\end{eqnarray*}
Here, the first term represents the standard action. The second term contains the c-number $\m{E}_{s,f}$ corresponding to the flux due to the  smooth background metric and leads to a non-degenerate (kinematical) vacuum state. Rather remarkably, it was shown in \cite{s} that the SU(2) Gauss law and the spatial diffeomorphism constraints can be unitarily implemented in the quantum theory based on this representation\footnote{The introduction of the background in the quantum kinematics can also be understood as an extension of the original set of operators, the new element being the `background exponential' operators, as noticed by Varadarajan\cite{m}}. Also the resulting spectra of the geometrical operators like area and volume have a simple additive structure, comprising the usual (LQG) contribution and the one depending solely on the background\cite{s}. 

Here, our interest in this representation lies in its applicability to the simpler context of two-dimensional PFT, and is inspired by the fact that this serves as a solvable toy model for quantum gravity. In particular, we set up a polymer quantization framework for this model with a non-degenerate background metric. We find that there are important differences as compared to the polymer quantized PFT without a background. The results here provide some important hints as to what the implications of this representation could be in the case of four-dimensional gravity theory. 

In the next section, we briefly review the classical PFT (see \cite{kuchar} for details). In section-III, we construct a polymer representation for this theory with a background. This includes a discussion of the gauge transformations, Dirac observables and conformal isometries. Next, we employ the group-averaging procedure to solve the constraints and discuss the essential features of the resulting space of physical states. In section-IV, we discuss the spacetime interpretation of the physical states and find the spectrum of the length operator. The final section contains some relevant remarks.
\section{Classical theory}
\subsection{Canonical formulation}
Based on the action (\ref{s-pft}), a canonical formulation for PFT can be developed\cite{kuchar}. The one dimensional spatial manifold  is chosen to be a circle. Such a choice of topology makes this system very similar to canonical gravity in four dimensions, where the three dimensional spatial slices are usually chosen to be compact. Thus, in PFT the initial Cauchy slice is given by the circle at $t=0$ with the angular coordinate $x\in [0,2\pi]$, where the points $x=0$ and $x=2\pi$ are identified. Each $t=$ constant circle defines a spacelike slice in the flat spacetime. The basic fields are the embedding variables $X^A(x^\alpha)$ and matter fields $\phi(x^\alpha)$. Their conjugate momenta are $\Pi_A(x^\alpha)$ and $\pi(x^\alpha)$, respectively. Thus, the basic Poisson brackets are given by:
\begin{eqnarray*}
\left\{ X^A(x),\Pi_A(y)\right\}~=~\delta(x,y),~\left\{\phi(x),\pi(y)\right\}~=~\delta(x,y) 
\end{eqnarray*}
where, $\delta(x,y)$ is the Dirac-delta function on the $x$-circle. A convenient way to analyse this theory is to define a new set of variables, corresponding to the mutually exclusive sectors of left and right moving fields. These are given by $(X^\pm(x),\Pi_\pm(x))$ for embedding sector, and $Y^\pm(x)$ for matter sector, and are defined as:
\begin{eqnarray*}
&&X^\pm(x)~=~T(x)\pm X(x),~~\Pi_\pm(x)~=~\frac{1}{2}\left[\Pi_0(x) \pm \Pi_1(x)\right]\\
&& Y^\pm~=~\pi(x)\pm \phi'(x)
\end{eqnarray*}
These satisfy the following Poisson-brackets:
\begin{eqnarray*}
&&\left\{ X^\pm(x),\Pi_\pm(y)\right\}~=~\delta(x,y),~~\left\{ X^\mp(x),\Pi_\pm(y)\right\}~=~0,\\
&&\left\{ Y^\pm(x),Y^\pm(y)\right\}~=~\pm [\del_x\delta(x,y)-\del_y\delta(x,y)],~~\left\{ Y^\mp(x),Y^\pm(y)\right\}~=~0
\end{eqnarray*}
As a consequence of the cylindrical topology of spacetime, the embedding variables satisfy quasiperiodic boundary conditions:
\begin{eqnarray}\label{qbc}
X^{\pm}(x+2\pi)~=~X^{\pm}(x)\pm 2\pi L
\end{eqnarray}
where, we have chosen $2\pi L$ as the length of the $T~$= constant slice in the inertial spacetime. 
All the other fields are periodic in $x$ with a periodicity of $2\pi$.

The general covariance of the action (\ref{s-pft}) gets reflected through the presence of two first-class constraints, given by:
\begin{eqnarray*}
H_\pm~=~\Pi_\pm(x) X^{\pm'}(x)\pm\frac{1}{4}Y^\pm(x)^2
\end{eqnarray*}
These obey the following algebra:
\begin{eqnarray*}
\left\{\int dx N^\pm(x) H_\pm(x),\int dx M^{\pm}(y)H_{\pm}(y)\right\}~&=&~\int dx [N^\pm (x)M^{\pm '}(x)-M^\pm(x)N^{\pm '}(x)] H_\pm(x)\\
\left\{\int dx N^\mp(x) H_\mp(x),\int dx M^\pm(y)H_\pm(y)\right\}~&=&~0
\end{eqnarray*}
where, $N^\pm,M^\pm$ are Lagrange multipliers. As discussed in\cite{laddha,laddha1}, the action of these constraints can be interpreted as two independent spatial diffeomorphisms provided one extends the domain of the fields from $[0,2\pi]$ to the entire real line. Such a description is consistent with the boundary conditions provided the embedding variable $X^\pm(x)$ has a quasiperiodic extension, while all the other fields $F^\pm(x)=(\Pi_\pm(x),Y^\pm(x))$ alongwith $N^\pm$ are extended periodically on the real line:
\begin{eqnarray*}
X^\pm_{ext}(x+2m\pi)~&=&~X^\pm(x)~\pm~2m\pi L,\\
F^\pm_{ext}(x+2m\pi)~&=&~F^\pm(x),\\
N^\pm_{ext}(x+2m\pi)~&=&~N^\pm(x)~~~\mathrm{for}~~ x\in[0,2\pi)
\end{eqnarray*}
Thus, under a finite gauge transformation characterised by a pair of periodic diffeomorphisms $(\phi^+,\phi^-)$, the fields transform as:
\begin{eqnarray*}
\left(\alpha_{\phi^\pm}X^\pm\right)(x)~&=&~X^\pm_{ext}(\phi^\pm(x))\\
\left(\alpha_{\phi^\mp}X^\pm\right)(x)~&=&~X^\pm(x)\\
\left(\alpha_{\phi^\pm}F^\pm\right)(x)~&=&~F^\pm_{ext}(\phi^\pm(x))\\
\left(\alpha_{\phi^\mp}F^\pm\right)(x)~&=&~F^\pm(x)
\end{eqnarray*}
%
\subsection{Dirac observables}
Integral of any periodic scalar density of weight one, which can be written in terms of the phase-space variables separately in the `$+$' or `$-$' sector, is a Dirac observable. 
We consider two particular examples of these below.

(i) Mode functions:

For  any real periodic real function $f^{\pm}(X^\pm)$, one can construct the observable:
\begin{eqnarray*}
O_f^{\pm}~=~\int_{S^1} dx~ Y^{\pm}(x) f^{\pm}(X^\pm(x))
\end{eqnarray*}
These essentially correspond to the Fourier modes of the free scalar field, which are given by:
\begin{eqnarray*}
a_n^{\pm}~=~\int_{S^1} dx ~Y^{\pm}(x) e^{i n X^{\pm}},~~n\in Z,~~n>0
\end{eqnarray*}

(ii) Conformal isometries:
%

The generators of the conformal isometries in PFT are also Dirac observables. These are given by:
\begin{eqnarray*}
O_c^{\pm}~=~\int_{S^1} dx~ \Pi_{\pm}(x) U^{\pm}(X^{\pm}(x))
\end{eqnarray*}
where, $U^{\pm}(X^{\pm}(x))$ is any periodic functional of $X^{\pm}(x)$. 
The action of these on the embedding $X^\pm(x)$ leads to a conformal scaling of the metric which leaves the PFT action invariant. The isometry generators satisfy the following algebra:
\begin{eqnarray*}
\left\{\int_{S^1} dx ~\Pi_{\pm}(x) U^{\pm}(X^{\pm}(x)),\int_{S^1} dy~ \Pi_{\pm}(y) V^{\pm}(X^{\pm}(y))\right\}=\int_{S^1} dx~ \Pi_{\pm}(x) (U^{\pm}V^{\pm'}-V^\pm U^{\pm'})(x)
\end{eqnarray*}
The finite action of these can be represented  through a pair of functions $\phi_c^{\pm}$ which act only on the embedding sector:
\begin{eqnarray*}
&&\alpha_{\phi^{\pm}_c}  X^{\pm}(x)=\phi^{\pm}_c(X^{\pm}(x)),~~\alpha_{\phi^{\mp}_c}  X^{\pm}(x)=X^{\pm}(x)),\\
&&\alpha_{\phi^{\pm}_c}  Y^{\pm}(x)= Y^{\pm}(x),~~\alpha_{\phi^{\mp}_c}  Y^{\pm}(x)= Y^{\pm}(x)
\end{eqnarray*}
The functions $\phi_c^\pm(X^\pm)$ are invertible, connected to identity, and monotonically increasing:
\begin{eqnarray*}
\frac{d\phi_{c}^\pm}{dX^\pm}>0
\end{eqnarray*}
Also, the boundary conditions (\ref{qbc}) imply the following condition on $\phi_c^\pm$:
\begin{eqnarray*}
\phi_c^\pm(X^\pm (x)+2n\pi L)~=~\phi_c^\pm(X^\pm (x))\pm 2 n \pi L
\end{eqnarray*}
 This ends the brief discussion of classical PFT. In what follows next, we set up a polymer quantization of PFT based on the new kinematical representation with a background, as originally introduced in \cite{k}, and study the implications. 
%
\section{Polymer quantization with a background}
We construct the kinematical Hilbert space based on the charge-network states separately for the embedding and matter sectors along the lines of ref.\cite{laddha1}. The new ingredient which enters the construction here is a nondegenerate background embedding, which affects the action of the basic operators on the states. 
\subsection{Embedding sector}
Let us first define a pair of graphs $\gamma_e^{\pm}$ each of which are composed of a finite number of non-overlapping (except at the vertices) edges, and covers the entire range of $x\in[0,2\pi]$. The embedding charge network $\ket{s_e^{\pm},\m{X^{\pm}}}:=\ket{\gamma_e^{\pm},\overrightarrow{k^{\pm}},\m{X}^{\pm}}$ is labelled by $\gamma_e^{\pm}$, a set of embedding charges $\overrightarrow{k^{\pm}}=(k_1^{\pm},..,k_{N^{\pm}}^{\pm})$ and the smooth background (embedding) charge $\m{X}^{\pm}(x)$. The total number of edges in $\gamma_e^{\pm}$ are $N^{\pm}$, and the $i$-th edge $e_i^{\pm}$ has an embedding charge $k_i^{\pm}=\alpha n^{\pm},~n^{\pm}\in Z$. The fixed real number $\alpha$ is a free parameter  in the theory and is an analogue of the Barbero-Immirzi parameter in LQG. The inner product between any two charge-nets is given by:
\begin{eqnarray*}
\braket{s^{'\pm}_e,\m{X}^{'\pm}|s^{\pm}_e,\m{X}^{\pm}}=\delta_{s^{'\pm}_e,s_e^{\pm}} ~\delta_{\m{X}^{'\pm},\m{X}^{\pm}}
 \end{eqnarray*}
 The basic operators are the embedding fields $\hat{X}^{\pm}(x)$ and the holonomy $\hat{h}_{\gamma_e^{\pm}}$ on the graph $\gamma_e^{\pm}$ defined as:
\begin{eqnarray}
 \hat{h}_{\gamma_e^{\pm}}~=~e^{i\sum_{i=1}^{N^{\pm}} k^{\pm}_{i} \int_{e_i\in \gamma_e} \Pi_{\pm}}
 \end{eqnarray}
Their action on the charge-network states are given below:
\begin{eqnarray*}
\hat{X}^{\pm}(x)\ket{\gamma_e^{\pm},\overrightarrow{k^{\pm}},\m{X}^{\pm}}&=&(k^{\pm}_{x,s_e^{\pm}}+\m{X}^{\pm}(x))\ket{\gamma_e^{\pm},\overrightarrow{k^{\pm}},\m{X}^{\pm}}\\
 \hat{h}_{\gamma_{e}^{'\pm}}\ket{\gamma_e^{\pm},\overrightarrow{k^{\pm}},\m{X}^{\pm}}&=&\ket{\gamma_e^{'\pm} \circ\gamma_e^{\pm},\overrightarrow{k^{'\pm}}+\overrightarrow{k^\pm},\m{X}^{\pm}}
\end{eqnarray*} 
where \begin{eqnarray*}
k^{\pm}_{x,s_e^{\pm}}&=&k^{\pm}_m \mathrm{~~~~~if ~x \in Interior(e^{\pm}_m)},\\
&=&\frac{1}{2}(k^{\pm}_m+k^{\pm}_{m+1}) \mathrm{~~~if~x \in e^{\pm}_m \cap e^{\pm}_{m+1}},\\
&=&\frac{1}{2}(k^{\pm}_1+k^{\pm}_{N^{\pm}}+\m{\lambda}^{\pm}) \mathrm{~~~if~x=0~or~2\pi}
\end{eqnarray*}
where, $\m{\lambda}^{\pm}$ are two arbitrary constants of the form $\m{\lambda}^{\pm}=\alpha m^{\pm}, ~m^{\pm} \in Z$.
At this stage, let us make a few relevant remarks below:\\
a) The fixed numbers $\m{\lambda}^{\pm}$ can be absorbed in the c-numbers $\m{X}^{\pm}$ and hence would be set to zero in the rest of our analysis;\\
b) In order to satisfy the quasiperiodic boundary condition on $\hat{X}^{\pm}(x)$, we take the $k_i^{\pm}$-s to be periodic and the background charges to be quasiperiodic, i.e., $k^{\pm}_{i+N^{\pm}}=k_i^{\pm}$, $\m{X}^{\pm}(2\pi)=\m{X}^{\pm}(0)\pm 2\pi L$; \\
c) Any two sets of embedding data $(k_i^{\pm},\m{X}^{\pm})$ and $(k_i^{\pm}+n^{\pm},\m{X}^{\pm}-n^{\pm})$ are identified to be the same data, where $n^\pm \in Z$.

\subsection{Matter sector:}
The basic operators corresponding to the matter fields $Y^{\pm}(x)$ are given by the matter holonomies  $\hat{h}_{\gamma_m^{\pm}}$ defined on the graphs $\gamma_m^{\pm}$:
\begin{eqnarray}\label{holonomy-m}
\hat{h}_{\gamma_m^{\pm}}=e^{i\sum_{i=1}^{N^{\pm}} l^{\pm}_{i} \int_{e_i \in \gamma_m} Y^{\pm}}\end{eqnarray}
Each edge $e_i^{\pm}$ carries a matter charge $l^{\pm}_i$. These charges are of the form $l^{\pm}_i=\beta n^{\pm}+c_0$, where $\beta$ is a real constant of dimension $h^{-\frac{1}{2}}$, $n$ are integers and $c_0 \in R$ corresponds to the zero mode of the scalar field\footnote{A detailed discussion of the scalar zero mode is not relevant for our purpose, and hence would not be attempted here. See \cite{laddha,laddha1} for further details.}. The charge network states for the matter sector $\ket{s_m^{\pm}}:=\ket{\gamma_m^{\pm},\overrightarrow{l^{\pm}}}$ are labelled by $\gamma_m^{\pm}$ and the set $\overrightarrow{l^{\pm}}=(l_1^{\pm},..,l^{\pm}_N)$. These provide a representation of the matter-holonomy operators as below:
 \begin{eqnarray*}\hat{h}_{\gamma_m^{'\pm}} \ket{\gamma_m^{\pm},\overrightarrow{l^{\pm}}}=e^{\frac{i}{2}\alpha({\gamma'}_m^{\pm},
 \gamma_m^{\pm})}\ket{\gamma_m^{'\pm}\circ\gamma_m^{\pm},~\overrightarrow{l^{'\pm}}+\overrightarrow{l^{\pm}}}
\end{eqnarray*} 
Here we have defined $\gamma_m^{'\pm}\circ\gamma_m^{\pm}$ as a graph finer than both $\gamma_m^{'\pm}$ and $\gamma_m^{\pm}$. The edges of $\gamma_m^{'\pm}\circ\gamma_m^{\pm}$ are characterised by the following distribution of matter charges: \\
(a) In regions (of $\gamma^{'\pm}_{m}\circ  \gamma^{\pm}_{m}$) where the edges of $\gamma^{'\pm}_{m}$ and $\gamma^{\pm}_{m}$ do not overlap, the charges are the same as the original ones;\\
(b) In regions where the edges overlap, the charge is a sum of the original charges.\\
Note that phase factor in the above equation arises due to the non-commutativity of the $Y^{\pm}(x)$'s.
\subsection{Nondegeneracy condition}
Classically, demanding the nondegeneracy of the spatial metric implies $\pm X^{\pm'}(x)>0$. In the quantum theory, this condition translates to
\begin{eqnarray*}
\pm\bra{\gamma_{e}^{\pm},\overrightarrow{k^{\pm}},\mathring{X}^{\pm}} \hat{X}^{\pm'}(x)\ket{\gamma_{e}^{\pm},\overrightarrow{k^{\pm}},\mathring{X}^{\pm}}\geqslant 0~
\end{eqnarray*} 
where, the background $\m{X}^{\pm}(x)$ itself is non-degenerate and smooth.
This implies:
\begin{eqnarray}\label{nondeg}
&&(a)~\pm(k_{m^{\pm}+1}^{\pm}-k_{m^{\pm}}^{\pm}) \geqslant 0 \mathrm{~for ~x \in e_{m^{\pm}}^{\pm} \cap e_{m^{\pm} +1}^{\pm}~,~1\leqslant m^{\pm}\leqslant (N^{\pm} -1)~~~and} \nonumber\\
&&(b)~\pm(k_{N^{\pm}}^{\pm}-k_1^{\pm})\geqslant 2\pi 
\end{eqnarray} 
Notice that these conditions are algebraically independent of the background.

\subsection{Unitary action of gauge transformations}
In the full (embedding plus matter) sector of the theory, we denote a typical charge-network state by $\ket{s_e^{\pm},\m{X}^{\pm}}\otimes \ket{s_m^{\pm}}$. On these, the action of the pair of gauge transformations $(\phi^+,\phi^-)$ are represented through the pair of unitary operators $(\hat{U}_{\phi^+},\hat{U}_{\phi^-})$:
\begin{eqnarray}
 \hat{U}_{\phi^{\pm}} \ket{s_e^{\pm},\m{X}^{\pm}}\otimes \ket{s_m^{\pm}}~=~\ket{\phi^{\pm}(s_e^{\pm}),\phi_*^{\pm} \m{X}^{\pm}}\otimes \ket{\phi^{\pm}(s_m^{\pm})}
 \end{eqnarray}
 where $\phi^{\pm}(s^{\pm})$ are the images of $s^{\pm}$ under the actions of $\hat{U}_{\phi^{\pm}}$ and $\phi^{\pm}_* \m{X}^{\pm}$ denotes the push-forward of $\m{X}^{\pm}$. The embedding charges $k^{\pm}_{x,\phi^{\pm}(s_e^{\pm})}$ in the new charge-network are related to the original ones as:
 \begin{eqnarray}
 k^{\pm}_{x,\phi^{\pm}(s_e^{\pm})}~=~k^{\pm}_{\phi^{{\pm}^{-1}}(x),s_e^{\pm}}
 \end{eqnarray}
 To see that the above action is indeed a representation, we note that 
 \begin{eqnarray*}
 && \hat{U}_{\phi_1^{\pm}} \hat{U}_{\phi_2^{\pm}} \ket{s_e^{\pm},\m{X}^{\pm}}\otimes \ket{s_m^{\pm}}\\
  &=& \hat{U}_{\phi_1^{\pm}}\ket{\phi_2^{\pm}(s_e^{\pm}),\phi_{2*}^{\pm} \m{X}^{\pm}}\otimes \ket{\phi_2^{\pm}(s_m^{\pm})}\\
  &=& \ket{\phi_1^{\pm}(\phi_2^{\pm}(s_e^{\pm})),(\phi_{1*}^{\pm} o \phi_{2*}^{\pm}) \m{X}^{\pm}}\otimes \ket{\phi_1^{\pm}(\phi_2^{\pm}(s_m^{\pm}))}\\
  &=& \ket{(\phi_1^{\pm} \circ\phi_2^{\pm})(s_e^{\pm}),(\phi_1^{\pm} \circ\phi_2^{\pm})_* \m{X}^{\pm}}\otimes \ket{(\phi_1^{\pm} \circ\phi_2^{\pm})(s_m^{\pm})}\\
  &=&\hat{U}_{(\phi_1^{\pm} \circ\phi_2^{\pm})} \ket{s_e^{\pm},\m{X}^{\pm}}\otimes\ket{ s_m^{\pm}}
 \end{eqnarray*}
 Also, it is easy to check the unitarity of $\hat{U}_{\phi^{\pm}}$ using the inner product:
\begin{eqnarray*}
&&\bra{s_m^{'\pm}}\otimes \bra{s_e^{'\pm},\m{X}^{'\pm}}\hat{U}^{\dagger}_{\phi^{\pm}} \hat{U}_{\phi^{\pm}} \ket{s_e^{\pm},\m{X}^{\pm}}\otimes \ket{s_m^{\pm}}\\
&=&  \bra{\phi^{\pm}(s_m^{'\pm})}\otimes \bra{\phi^{\pm}(s^{'\pm}_e),\phi^{\pm}_* \m{X}^{'\pm}}\hat{I}\ket{\phi^{\pm}(s_e^{\pm}),\phi_*^{\pm} \m{X}}\otimes\ket{\phi^{\pm}(s_m^{\pm})}\\
&=& \delta_{\phi^{\pm}(s_e^{\pm}),\phi(s^{'\pm}_e)}~\delta_{\phi^{\pm}_* \m{X}^{\pm},\phi^{\pm}_* \m{X}^{'\pm}}~\delta_{\phi^{\pm}(s_m^{\pm}),\phi^{\pm}(s^{'\pm}_m)}\\
&=& \delta_{s_e^{\pm},s_e^{'\pm}}~\delta_{\m{X}^{\pm},\m{X}^{'\pm}}~\delta_{s_m^{\pm},s_m^{'\pm}}
\end{eqnarray*}
where we have defined $\hat{I}$ as the identity operator in the first line and have used the invertibility of $\hat{U}_{\phi^{\pm}}$ in the last line.

The representation above induces the correct transformations of the basic operators. We demonstrate this explicitly for $\hat{X}^{\pm}(x)$ below:
\begin{eqnarray*}
&& \hat{U}_{\phi^{\pm}}\hat{X}^{\pm}(x) \hat{U}^{\dagger}_{\phi^{\pm}}\ket{s_e^{\pm},\m{X}^{\pm}}\otimes \ket{s_m}\\
 &=& \hat{U}_{\phi^{\pm}} \hat{X}^{\pm}(x)\ket{\phi^{\pm^{-1}}(s_e^{\pm}),\phi^{\pm^{-1}}_* \m{X}^{\pm}}\otimes \ket{\phi^{\pm^{-1}}(s_m^{\pm})}\\
 &=& \hat{U}_{\phi^{\pm}}\left[k^{\pm}_{x,\phi^{\pm^{-1}}(s_e)}+\m{X}^{\pm}(\phi^{\pm}(x))\right]\ket{\phi^{\pm^{-1}}(s_e^{\pm}),\phi^{\pm^{-1}}_* \m{X}^{\pm}}\otimes \ket{\phi^{\pm^{-1}}(s_m^{\pm})}\\
 &=& \left[k^{\pm}_{\phi^{\pm}(x),s_e}+\m{X}^{\pm}(\phi^{\pm}(x))\right]\ket{s_e^{\pm},\m{X}^{\pm}}\otimes \ket{s_m^{\pm}}\\
 &=& \hat{X}^{\pm}(\phi^{\pm}(x)) \ket{s_e^{\pm},\m{X}^{\pm}}\otimes \ket{s_m^{\pm}}
\end{eqnarray*}
where in the third line we have used the identity $k^{\pm}_{x,\phi^{\pm^{-1}}(s^{\pm}_e)}=k^{\pm}_{\phi^{\pm}(x),s^{\pm}_e}$. Similarly it can be checked that
the other operators also transform correctly.
\subsection{Dirac observables corresponding to mode functions}
The classical Dirac observables are of the form $\int_{S^1} f^{\pm}(X^{\pm}(x)) Y^{\pm}(x)$, where $f^{\pm}(X^{\pm})$ are periodic functions of the 
embedding variable $X^{\pm}$. In the quantum theory, these observables as they are do not admit a well-defined representation. However, their exponentials 
do, and they can be represented as unitary operators on the kinematic Hilbert space. The action of these exponential operators on the kinematical states is given below:
\begin{eqnarray}
 &&\widehat{e^{i\int_{S^1} f^{\pm}(X^{\pm}(x)) Y^{\pm}(x)}}\ket{\gamma_e^{\pm},\overrightarrow{k^{\pm}},\m{X}^{\pm}}\otimes\ket{\gamma_m^{\pm},\overrightarrow{l^{\pm}}}\\
 &=&\widehat{e^{i\int_{S^1} f^{\pm}(X^{\pm}(x)) Y^{\pm}(x)}}\ket{\gamma_e^{\pm},\overrightarrow{k^{\pm}},\m{X}^{\pm}}\otimes\ket{\gamma_m^{\pm},\overrightarrow{l^{\pm}}}\nonumber\\
 &=& \widehat{e^{i\int_{S^1} f^{\pm}(k^{\pm}_{x,s_e^{\pm}}+\m{X}^{\pm}(x)) Y^{\pm}(x)}}\ket{\gamma_e^{\pm},\overrightarrow{k^{\pm}},\m{X}^{\pm}}\otimes\ket{\gamma_m^{\pm},\overrightarrow{l^{\pm}}}\nonumber\\
 &=& \hat{h}_{\gamma^{\pm}_{m_{f}}}~\ket{\gamma_e^{\pm},\overrightarrow{k^{\pm}},\m{X}^{\pm}} \otimes \ket{\gamma_m^{\pm},\overrightarrow{l^{\pm}}}\nonumber\\
 &=& e^{\frac{i}{2}\alpha(\gamma_{m_{f}}^{\pm},\gamma_m^{\pm})}\ket{\gamma_e^{\pm},\overrightarrow{k^{\pm}},\m{X}^{\pm}}\nonumber \otimes \ket{\gamma_{m_{f}}^{\pm} \circ \gamma_m^{\pm},~\overrightarrow{l^{\pm}}+\overrightarrow{f^{\pm}}(k^{\pm}+\m{X}^{\pm})}
\end{eqnarray}
where, the graph $\gamma_{m_{f}}^{\pm} o  \gamma_{m}^{\pm}$ is finer than both $\gamma_{m_{f}}^{\pm}$ and $\gamma_{m}^{\pm}$ and we define the matter holonomy operator $\hat{h}_{\gamma^{\pm}_{m_{f}}}$ as:
\begin{eqnarray} \label{h-f}
\hat{h}_{\gamma^{\pm}_{m_{f}}}=\widehat{e^{i\int_{S^1} f^{\pm}(k_{x,s}^{\pm}+\m{X}^{\pm} (x)) Y^{\pm}(x)}}
\end{eqnarray} 
Note that the definition (\ref{h-f}) is consistent with (\ref{holonomy-m}) provided the (periodic) function $f^{\pm}(X^{\pm})$ is piecewise constant, and satisfy the condition: $f^{\pm}(\alpha n^{\pm}+\m{X}^{\pm})=\beta n^{'\pm}+c_0$ with $n^{\pm},n^{'\pm} \in Z$, $c_0 \in R$. 
\subsection{Conformal isometries}
We represent the action of the conformal isometries $\phi_c:=(\phi^{+}_c,\phi^{-}_c)$ by a pair of unitary operators $\hat{V}_{\phi_c^{\pm}}$ on the states:
\begin{eqnarray}\label{v-phi}
 \hat{V}_{\phi_{c}^{\pm}}\ket{\gamma_e^{\pm},\overrightarrow{k},\m{X}^{\pm}}\otimes \ket{\gamma_m^{\pm},\overrightarrow{l}}~=~ \ket{\gamma^{\pm}_{e},\overrightarrow{\bar{k}},\m{\bar{X}}^{\pm}}\otimes \ket{\gamma_m^{\pm},\overrightarrow{l}}
\end{eqnarray}
where, at any point $x$ in the graph $\gamma_e$, the embedding charges $\overrightarrow{(\bar{k}}^{\pm},\m{\bar{X}}^{\pm})$ in the new charge-net $\ket{\gamma^{\pm}_{e},\overrightarrow{\bar{k}},\m{\bar{X}}^{\pm}}$ are related to the original ones  through the mapping $\phi_c^{\pm}(X^{\pm})$ as:
\begin{eqnarray*}
 \bar{k}^{\pm}+~\m{\bar{X}}^{\pm}=\phi_c^{\pm^{-1}}(k^{\pm}+\m{X}^{\pm})
\end{eqnarray*}
 Note that the above representation leads to the correct transformation of the operators $X^{\pm}(x)$:
\begin{eqnarray*}
\hat{V}_{\phi_c^{\pm}} \hat{X}^{\pm}(x)\hat{V}^{\dagger}_{\phi_c^{\pm}}\ket{\gamma_e^{\pm},\overrightarrow{k},\m{X}^{\pm}}\otimes \ket{\gamma_m^{\pm},\overrightarrow{l}}~&=&~\phi^{\pm}_c(k^{\pm}_{x,s_e^{\pm}}+\m{X}(x))\ket{\gamma_e^{\pm},\overrightarrow{k},\m{X}^{\pm}}\otimes \ket{\gamma_m^{\pm},\overrightarrow{l}}\\
~&=&~\phi^{\pm}_c(\hat{X}(x))\ket{\gamma_e^{\pm},\overrightarrow{k},\m{X}^{\pm}}\otimes \ket{\gamma_m^{\pm},\overrightarrow{l}}
\end{eqnarray*}

The map $\phi_c^{\pm}$ is invertible and monotonic, and is such that the profile of the total embedding charge
$(k^{\pm}+\m{X}^{\pm})$, which 
 exhibit jumps across the vertices of the graph, gets mapped to a new profile 
 $(\bar{k}^{\pm}+\m{\bar{X}}^{\pm})$, which also has a similar discontinuous structure. To emphasize, the amount of the jump at a particular vertex is the same for the two profiles. At the $n$-th vertex, the jump is given by the difference $k^{\pm}_{n+1}-k^{\pm}_{n}=\alpha m^{\pm}_n$, where each $m^{\pm}_n$ is a fixed integer. It can be shown that non-linear functions in general cannot provide such a mapping (for a proof, see Appendix A). The only possible mappings are the linear ones, corresponding to constant real-valued translations. 
 Thus, in our case, the action of the operator $\hat{V}_{\phi_c}$ on a charge-net can be characterised by two constant real numbers $r^+,r^-$: 
 \begin{eqnarray}
 \phi_c^{\pm}(k^{\pm}_{x,s_e}+\m{X}^{\pm} (x))~=~k^{\pm}_{x,s_e}+\m{X}^{\pm} (x)+r^{\pm}
 \end{eqnarray}
 Since $r^{\pm}$ above can be any real number, the rigid translation can be of any arbitrary amount. Thus, not the whole conformal group, but only this abelian subgroup of continuous translations is implemented in the quantum theory. 
 
The features as described above are different compared to the case where there is no background\cite{laddha,laddha1}. There, for embedding charges of the form $k^{\pm}=\alpha n^{\pm},~\alpha\in R,~n^{\pm}\in Z$ as here, only the abelian subgroup of discrete translations are implemented in the quantum theory\cite{laddha1}. In contrast, for real-valued charges $k^{\pm}$, there are no such restrictions, and the whole conformal group can be represented in the quantum theory\cite{laddha}. Notice that in our case here with a nondegenerate background, we have an intermediate scenario since the (sub)group of continuous translations is larger than the one containing only discrete translations. Thus, the introduction of the smooth background $\m{X}^{\pm}$ results in an enlargement of the symmetry group as implemented in the quantum theory.

\subsection{Group averaging} 
To simplify the notation, we present the explicit computations only for the `$+$' sector and suppress the `$+$' indices from now on. The analysis for the `$-$' sector is exactly similar. Also, we assume that the distribution of the embedding charges in the states are consistent with the nondegeneracy condition (\ref{nondeg}), since this is physically relevant.

We apply the group averaging map (or $\eta$ map) to $\psi_{s_e,\m{X},s_m}=\ket{s_e,\m{X}}\otimes \ket{s_m}$, an element of the kinematic Hilbert space $H_{kin}$, in order to find a formal solution of the constraints in a suitable space $\Phi^*\supset H_{kin}$\cite{almt,marolf}:
\begin{eqnarray}\label{etamap}
 \bra{\eta[\psi_{s_e,\m{X},s_m}]}=\eta_{[s_e,\m{X},s_m]}\sum_{\phi \in G[s_e,\m{X},s_m]}~_{\phi}\bra{s_m}~\otimes~_{\phi}\bra{s_e,\m{X}} 
\end{eqnarray}
where, the positive real coefficient $\eta_{[s_e,\m{X},s_m]}$, depends only on the gauge orbit of $\ket{s_e,\m{X}}\otimes\ket{s_m}$, and the set of gauge transformations $G[s_e,\m{X},s_m]$ is such that the sum is only over distinct images of $\ket{s_e,\m{X}}\otimes\ket{s_m}$. The elements $\bra{\eta}\in \Phi^*$ are generalized states in the sense that they are not normalizable. These are linear functionals on the subspace $\Phi \subset H_{kin}$ of test functions, which are given by any finite linear combination of the charge-network states. As is obvious, $\bra{\eta}$ is a solution of the constraints.

Given this $\eta$-map, the inner product between the physical states is defined as:
\begin{eqnarray*}
 \braket{\eta[\psi_{s_e,\m{X},s_m}]|\eta[\psi_{s'_e,\m{X'},s'_m}]}_{phy}~=~\bra{\eta[\psi_{s_e,\m{X},s_m}]}[\psi_{s'_e,\m{X'},s'_m}]
\end{eqnarray*}

Next, in order to reduce the ambiguity in the coefficients $\eta_{[s_e,\m{X},s_m]}$, we use the commutativity of the $\eta$ map with any Dirac observable $\hat{O}$ in the theory:
\begin{eqnarray}\label{O-obs}
\bra{\eta[\psi]}\hat{O}^{+} ~=~\bra{\eta[\hat{O}\psi]}
\end{eqnarray}
For $\hat{O}=\widehat{e^{i\int_{S^1} f(X)Y}}$, the r.h.s. of eqn.(\ref{O-obs}) gives:
\begin{eqnarray}\label{rhs}
 \bra{\eta[\widehat{e^{i\int_{S^1} f(X)Y}}\psi_{s_e,\m{X},s_m}]}
 &=& e^{\frac{i}{2}\alpha(s_{m_{f}},s_m)}~\eta_{[s_e,\m{X},s_{m_f}]} \sum_{\phi \in G[s_e,\m{X},s_{m_f}]}~ _{\phi}\bra{s_{m_f}}~\otimes~ _{\phi}\bra{s_e,\m{X}}
\end{eqnarray}
Here we have defined $\bra{s_{m_f}}=\bra{\gamma_m,\overrightarrow{l}+\overrightarrow{f}(k+\m{X})}$.
The L.H.S. gives:
\begin{eqnarray}\label{lhs}
\eta[\psi_{s_e,\m{X},s_m}]\widehat{e^{-i\int_{S^1} f(X)Y}}&=&\eta_{[s_e,\m{X},s_m]} \sum_{\phi' \in G[s_e,\m{X},s_m]}~ _{\phi'}\bra{s_m}\otimes~ _{\phi'}\bra{s_e,\m{X}}
\widehat{e^{-i\int_{S^1} f(X)Y}}\nonumber\\
&=& e^{\frac{i}{2}\alpha(s_{m_{f}},s_m)}~ \eta_{[s_e,\m{X},s_m]} \sum_{\phi' \in G[s_e,\m{X},s_m]}~_{\phi'}\bra{s_{m_f}} \otimes ~_{\phi'}\bra{s_e,\m{X}}
\end{eqnarray}
In the last line we have used the commutation of $\hat{U}_{\phi}$ and $\widehat{e^{i\int_{S^1} f(X)Y}}$. 
Now, notice that $\hat{U}_{\phi}~e^{i\int_{S^1} f(\hat{X}(x))}~\psi_{s_e,\m{X},s_m}\neq e^{i\int_{S^1} f(\hat{X}(x))}~\psi_{s_e,\m{X},s_m}$ iff
 $\hat{U}_{\phi}\psi_{s_e,\m{X},s_m}\neq \psi_{s_e,\m{X},s_m}$. Thus, the same $\hat{U}_{\phi}$ can generate the orbits (or, equivalence classes) $[s_e,\m{X},s_{m_f}]$ and $[s_e,\m{X},s_m]$, and the sets of gauge transformations $G[s_e,\m{X},s_{m_f}]$ and $G[s_e,\m{X},s_m]$ can be chosen to be the same.
  Thus, comparing the equations (\ref{rhs}) and (\ref{lhs}), we obtain that 
 \begin{eqnarray*}
\eta_{[s_e,\m{X},s_{m}]}= \eta_{[s_e,\m{X},s_{m_f}]}
\end{eqnarray*}
Now, the function $f(X)$ can always be chosen to make the sum $\overrightarrow{l}+\overrightarrow{f}=0$, making all the matter charges vanish. Such a distribution of 
matter charges is gauge invariant. This implies that the $\eta$-coefficients do not depend on the matter labels, and can be denoted as $\eta_{[s_e,\m{X}]}$. 

One can do a similar analysis for the conformal isometries $\hat{V}_{\phi_c}$. However, this does not lead to a significant reduction in the ambiguity. This is so because given any two states $\ket{s_e,\m{X}}:=\ket{\gamma_e,\overrightarrow{k},\m{X}}$ and $\ket{s'_e,\m{X'}}:=\ket{\gamma_e,\overrightarrow{k'},\m{X'}}$, it is not possible to find a $\hat{V}_{\phi_c}$ (given by a constant rigid translation) such that $\hat{V}_{\phi_c}\ket{\gamma_e,\overrightarrow{k},\m{X}}=\ket{\gamma_e,\overrightarrow{k'},\m{X'}}$. A rigid translation can interpolate between the values of $\m{X}(x)$ and $\m{X'}(x)$ only at one single point $x=x_0$ at the most, and not at all points $x$. This only removes one real number worth of ambiguity in $\eta_{[s_e,\m{X}]}=\eta_{[\gamma_e,k,\m{X}]}$. 

It is worth noting that there exists a superselected sector in the quantum theory where $\eta_{[s_e,\m{X}]}$ does not depend on the $k$-labels. This is defined by the condition $k_{n+1}=k_n+\delta_0$ where $n=1,..,N-1$ and $\delta_0$ is a fixed number of the form $\delta_0=\alpha m,~m\in Z$. Within such a sector, one can always identify a given set of embedding data $(\overrightarrow{k},\m{X})$ as $(\overrightarrow{k'},\m{X'})$ (see section III-A) where,
\begin{eqnarray*}
\overrightarrow{k}&=& \left(k,~k+\delta_0,..,~k+[N-1]\delta_0 \right)\\
\overrightarrow{k'}&=& \left(k',~k'+\delta_0,..,~k'+[N-1]\delta_0 \right)\\
\m{X'}(x)&=& \m{X}(x)+k-k'
\end{eqnarray*}
Thus, within this sector, the coefficients $\eta_{[s_e,\m{X}]}$ can be denoted as $\eta_{[\gamma_e,\m{X}]}$.
\subsection{Nonseparability}
 A gauge invariant state (\ref{etamap}) is completely  characterised by the set ($\overrightarrow{k},\overrightarrow{l},\overrightarrow{\m{X}}(v)$), where $\m{X}(v)$ denotes the value of $\m{X}$ at the vertex $v$, and we define $\overrightarrow{\m{X}}(v)=(\m{X}(v_1),...,\m{X}(v_N))$ with $N$ being the total number of vertices. In other words, this set of gauge invariant data labels the basis states in the physical space (see Appendix B for the proof). In this set, both $\overrightarrow{k}$ and $\overrightarrow{l}$ are countable labels, parametrised by integers. However, the set of $N$ real numbers $\overrightarrow{\m{X}}(v)$ can vary continuously from one basis to another, and are uncountable. Thus, the basis states are uncountably infinite, leading to a nonseparable physical Hilbert-space. This is in contrast to the case without the background with integer-labelled embedding charges\cite{laddha1}, where the physical space within a particular superselected  sector is separable. Note that this space can be recovered by putting the background charges $\m{X}$ to be zero.

%

\section{Quantum geometry}
\subsection{Spacetime interpretation}
The gauge invariant data associated with a physical state completely determines the emergent picture of the inertial spacetime in the quantum theory. These data are given by the distribution of the total embedding and matter charges, which essentially specifies the way in which matter data sits on the flat spacetime. 

For an embedding charge-network state $\ket{\gamma_{e}^{\pm},\overrightarrow{k^{\pm}},\m{X}^{\pm}}$ with $N^{\pm}$ edges, the total embedding charge $k^{\pm}_{n,tot}=k_n^{\pm}+\m{X}^{\pm}$ for the $n$-th edge lies within the range [$k_n^{\pm}+\m{X}^{\pm}_{min},~k_n^{\pm}+\m{X}^{\pm}_{max}]=[k_n^{\pm}+
\m{X}^{\pm}(v_n),~k_n^{\pm}+\m{X}^{\pm}(v_{n+1})]$, where $\m{X}^{\pm}(v_n)$ is the value of $\m{X}^{\pm}$ at the $n$-th vertex $v_n$. Now notice that the values of $k^{\pm}_{n,tot}$ also denote the inertial spacetime coordinates along the (null) `$+$' and `$-$' directions. In particular, each continuous interval [$k_n^+ +\m{X}^+_{min},~k_n^+ +\m{X}^+_{max}$] represents a spacetime strip, finite along the `$+$' direction and infinite along the `$-$' direction. Each such strip is followed by a void given by the amount $k_{n+1}^+ -k_n^+=\alpha m^+$, $m^+\in Z$. A similar interpretation holds for the total embedding data in the `$-$' sector. Thus, each pair of intervals [$k_n^{+}+\m{X}^{+}_{min},~k_n^{+}+\m{X}^{+}_{max}]$ and [$k_n^{-}+\m{X}^{-}_{min},~k_n^{-}+\m{X}^{-}_{max}]$ corresponds to a rectangular strip of flat spacetime, which is the region of intersection between the two strips along the `+' and `$-$' directions. Now, a physical (group-averaged) state corresponds to all data obtained by  quasiperiodically extending the initial data $(k^{\pm}_{1,tot},...,k^{\pm}_{N^{\pm},tot})$. This set of values of the total embedding charge contains all possible values of the inertial coordinates, and generates the whole spacetime. Thus, the inertial spacetime is made up of discrete rectangular strips. The matter charges correspond to discrete points on these spacetime strips. 

The (quantum) geometry of the flat spacetime as above is remarkably different from the case without the background, where one obtains a spacetime lattice\cite{laddha1} with a characteristic minimum length. The absence of any smallest length scale in the case here implies that there is no obstruction in resolving spacetime points which are arbitrarily close to each other.

It is important to note that the discrete geometry of the quantum spacetime is tied up with the discontinuous nature of the profile of the total embedding charge. This supercedes the classical spacetime continuum. In the trivial case where all integer-labelled embeddings $k_i$ are chosen to be equal or zero, which also is consistent with the nondegeneracy condition (\ref{nondeg}), we recover a flat spacetime continuum, as expected.

\subsection{Length operator}
The classical expression of length for PFT is given by:
\begin{eqnarray*}
L~=~\int dx ~|X^{+'}(x) X^{-'}(x)|^{\frac{1}{2}}
\end{eqnarray*}

First we define a triangulation of the graph $\gamma_e$ into infinitesimally small intervals $I_m\in [x_m-\frac{\Delta_m}{2},x_m+\frac{\Delta_m}{2}]$ of length $\Delta_m$. Using this triangulation, the length can be written as a discrete sum:
\begin{eqnarray*}
&&\sum_m \lim_{\Delta_m \to 0} |\Delta_m| |X^{+'}(x_m)X^{-'}(x_m)|^{\frac{1}{2}}\\
&=&\sum_m \lim_{\Delta_m \to 0} |\Delta_m| \left|\left[\frac{X^{+}(x_m+\frac{\Delta_m}{2})-X^{+}(x_m-\frac{\Delta_m}{2})}{\Delta_m}\right]\left[\frac{X^{-}(x_m+\frac{\Delta_m}{2})-X^{-}(x_m-\frac{\Delta_m}{2})}{\Delta_m} \right]\right|^{\frac{1}{2}}
\end{eqnarray*}

This discretised expression can be represented as a well-defined operator in the quantum theory. Note that the operator acts only on the embedding sector. 

For the case where the point $x_m$ does not coincide with any of the vertices of $\gamma_e$, its action on a charge-network state gives:
\begin{eqnarray*}
&&\hat{L}\ket{\gamma_e,\overrightarrow{k},\m{X}}\\
&=&\sum_m \lim_{\Delta_m \to 0} |\Delta_m|\left|X^{+'}(x_m)X^{-'}(x_m)\right|^{\frac{1}{2}} \ket{\gamma_e,\overrightarrow{k},\m{X}}\\
&=&\sum_m \lim_{\Delta_m \to 0} \left|\left(\m{X}^{+}(x_m+\frac{\Delta_m}{2})-\m{X}^{+}(x_m-\frac{\Delta_m}{2})\right)\left(\m{X}^{-}(x_m+\frac{\Delta_m}{2})-\m{X}^{-}(x_m-\frac{\Delta_m}{2}) \right)\right|^{\frac{1}{2}}\ket{\gamma_e,\overrightarrow{k},\m{X}}\\
&=&\sum_m \lim_{\Delta_m \to 0} |\Delta_m|\left|\m{X}^{+'}(x_m)\m{X}^{-'}(x_m)\right|^{\frac{1}{2}}\ket{\gamma_e,\overrightarrow{k},\m{X}}\\
&=&\int dx ~|\m{X}^{+'}(x)\m{X}^{-'}(x)|^{\frac{1}{2}}\ket{\gamma_e,\overrightarrow{k},\m{X}}
\end{eqnarray*}
which is just the contribution from the background.

Next we consider the case when the point $x_m$ coincides with one of the vertices $v_m$. If there are N vertices, the corresponding action on the state can be written as:
\begin{eqnarray*}
&&\sum_{m=1}^{N} \lim_{\Delta_m \to 0} |\Delta_m|\left|X^{+'}(v_m) X^{-'}(v_m)\right|^{\frac{1}{2}}\ket{\gamma_e,\overrightarrow{k},\m{X}}\\
&=&\sum_{m=1}^{N} \lim_{\Delta_m \to 0}\left[k^{+}_{m+1} +\m{X}^{+}(v_m+\frac{\Delta_m}{2})-k^+_m -\m{X}^{+}(x_m-\frac{\Delta_m}{2})\right]^{\frac{1}{2}}\\
&~&~~~~~~~\times~\left[k^-_{m}+\m{X}^{-}(x_m - \frac{\Delta_m}{2})- k^-_{m+1}-\m{X}^{-}(x_m+\frac{\Delta_m}{2})\right]^{\frac{1}{2}}\ket{\gamma_e,\overrightarrow{k},\m{X}}\\
&=&\sum_{m=1}^{N}\left[(k^+_{m+1}-k^+_{m})(k^-_{m}-k^-_{m+1})\right]^{\frac{1}{2}} \ket{\gamma_e,\overrightarrow{k},\m{X}}
\end{eqnarray*}

Thus, the total contribution becomes:
\begin{eqnarray}\label{spec}
 \hat{L}\ket{\gamma_e,\overrightarrow{k},\m{X}}~=~\left(\int dx ~|\m{X}^{+'}(x)\m{X}^{-'}(x)|^{\frac{1}{2}}
 +\sum_{m=1}^{N}\left[(k^+_{m+1}-k^+_{m})(k^-_{m}-k^-_{m+1})\right]^{\frac{1}{2}}\right) \ket{\gamma_e,\overrightarrow{k},\m{X}}\nonumber\\
\end{eqnarray}
Thus, the eigenvalue splits into background dependent and independent parts. This is remarkably similar to the general structure of higher dimensional geometrical operators like area and volume in kinematical LQG with a background, as discussed in \cite{s}. Due to the presence of the first term in equation (\ref{spec}), the spectrum is continuous, in contrast to the case without the background\cite{laddha2}. Thus, the quantum geometry as obtained here is not associated with an analogue of the Planck length, i.e. the finest spacing between two spacetime points. 
\section{Concluding remarks}
Here we have constructed a quantum theory of a parametrized scalar field on a cylindrical spacetime, based on a polymer representation with a nondegenerate vacuum. In other words, the geometry of the vacuum state is characterised by a nondegenerate spatial metric, unlike the standard polymer representation. The representation of this kind was originally introduced in the context of Loop Quantum Gravity\cite{k,s}, with a view to incorporate the notion of a smooth effective spacetime within the quantum kinematics. However, since a complete solution of the Hamiltonian constraint of gravity within LQG is still elusive, it is difficult to understand how this new representation affects the full theory, including its dynamical content. In contrast, the much simpler theory of a parametrised scalar field, which serves as a solvable toy model of canonical quantum gravity, provides a perfect arena where such a kinematical construction can be applied and the consequences can be analysed in detail. This is precisely the perspective from which a polymer quantization of two-dimensional PFT with a background has been set up here. 

%
The invariance of the PFT action under spacetime diffeomorphisms gets reflected through the presence of a pair of first-class constraints in the canonical theory. These generate motions of both embedding and matter data, and hence encode the dynamics. The finite action corresponding to these can be represented unitarily in the quantum theory. Classical PFT is also characterised by the symmetry group of conformal isometries, which generates arbitrary scalings of the metric while leaving the action invariant. However, in the quantum theory, only the group of (global) continuous translations can be implemented. Although this is a small abelian subgroup of the whole conformal group, this is still larger than the group of discrete translations as in the case without the background (with integer-labelled embedding charges)\cite{laddha1}. In this sense, the introduction of the background leads to an enlargement of the symmetry group as realized in the quantum theory.

In order to construct a physical space of states, we solve the constraints using the group-averaging procedure. The resulting physical space has an uncountable number of basis states, and hence is nonseparable. For a qualitative comparison, let us recall the cases without the background, where (within a suitable superselected sector) the physical state space is separable for integer-labelled charges\cite{laddha1}, and nonseparable for real-valued charges\cite{laddha}. One might think that since the physical space with the background is nonseparable anyway, we might as well work within the formulation with real-valued embedding and matter labels without introducing a background. From our viewpoint, this is not too appealing though, since in that case, the explicit similarity between the spin-networks in LQG, whose edges have discrete labels, and the charge-networks would be lost, and there would not be a straightforward way to relate to the notion of a smooth classical geometry within the quantum kinematics.

The gauge invariant data carried by the physical states completely characterises the geometry of the physical inertial spacetime, which is shown to be made up of discrete strips. This replaces the smooth classical geometry. In the `quantum' spacetime, the regions within a strip define a continuum, whereas the voids between the strips lead to discreteness. Each matter data (charge) can be associated with a particular spacetime strip. Again, this is unlike the cases without background, where the real and integer-labelled embedding charges correspond to continuous and discrete (lattice) geometries of the flat spacetime, respectively\cite{laddha,laddha1}. We emphasize that the discreteness in our case with a smooth background arises due the integer-valuedness of the embedding charges. 

Finally, we analyse the length operator and its eigenspectrum within this framework. We find that the eigenvalue splits into two parts. While one depends only on the background charge, the other contains only the integer embedding charges. The full spectrum is continuous. This is exactly the structure shared by the area and volume operators in LQG kinematics with a background\cite{s}. Thus, such a decoupling of the spectra of geometrical operators emerges as a robust feature of the representation with a nondegenerate vacuum.

We emphasize that the kinematical representation with a smooth background can be expected to be particularly relevant in the context of loop quantization of gravity in asymptotically flat spacetimes. In contrast to the usual LQG representation, the states here have continuous background labels which can be used to capture the smoothness of the spacetime geometry at asymptotia. As a preparation for a detailed analysis along these lines, which would be reported elsewhere, we have studied the implications of the new kinematic representation in the simpler context of compact PFT in this paper. It is our hope that the insights gained here can be used to construct a canonical quantization of gravity in asymptotically flat spacetimes.
%
%
%

\acknowledgments
It is a pleasure to thank M. Varadarajan for suggesting this problem, and for his encouragement. The author is also grateful to M. Campiglia and A. Laddha for many enlightening discussions, and to G. Date and H. Sahlmann for their comments on the manuscript.

\appendix
\section{Proof-1 (Conformal isometries)}

 Let us denote the total embedding charge as $X(x)$ at any point $x$. From  (\ref{v-phi}) it is evident that $\phi_c$ maps $X=k+\m{X}$ to $\bar{X}=\bar{k}+\m{\bar{X}}$. Now, let us take $\phi_c(X)$ to be a non-linear map given by $\phi_c(X)=X^n$, where $n$ is any integer. In general, one can also take any linear combination of functions of such form. Now, consider the values of this function across a vertex v, where the total embedding charges $X$ and $\bar{X}$ are discontinuous. It is crucial to note that the jumps in these two profiles at the particular vertex $v$ are exactly of the same amount $\Delta_n=\alpha n,~n\in Z$, where $\alpha$ is a fixed real number as defined earlier. While our analysis below applies to any value of $\alpha$, here we shall choose $\alpha=1$ for convenience. With this choice, the jumps across the vertices become integer-valued.
 
 The total embedding charge $X(x)$ at an infinitesimal distance $\epsilon$ away at the left and right of a vertex $v$ is given by $X(v-\epsilon)=k+X^0(v-\epsilon)$ and $X(v+\epsilon)=k'+X^0(v+\epsilon)$, respectively, where $k$ and $k'$ are two different integers. Under $\phi_c$, these total charges are mapped to  $\phi_c(X(v+\epsilon))$ and
 $\phi_c(X(v-\epsilon))$, respectively. Using the binomial expansion and then taking the limit $\epsilon\rightarrow 0$, the difference between these can be written as:
 \begin{eqnarray*}
 &&\lim_{\epsilon\rightarrow 0} \left[\phi_c(X(v-\epsilon))~-~\phi_c(X(v-\epsilon))\right]\\
 &=& \lim_{\epsilon\rightarrow 0}\left[\left(k'+\m{X}(v+\epsilon)\right)^n ~-~\left(k+\m{X}(v-\epsilon)\right)^n \right]\\
 &=& (k'^n-k^n)+c_1(k'^{n-1}-k^{n-1})\m{X}(v)+c_2(k'^{n-2}-k^{n-2})\m{X}(v)^2+...+c_{n-1}(k'-k)\m{X}(v)^{n-1}
 \end{eqnarray*}
 where, the $c_n$-s are the binomial coefficients.
 Now notice that only the first term is an integer, whereas for any general $\m{X}\in R$, the rest are all real-valued. Thus, although the jump across $v$ was integer-valued to begin with, under a general non-linear mapping $\phi_c$ and for any general $\m{X}$, the jumps would not be integer-valued anymore. Hence, the only allowed mappings are the linear ones, i.e., constant real-valued translations.  
\section{Proof-2 (Gauge-invariant labels)}
Let us consider a physical state given by:
\begin{eqnarray}
 \bra{\eta[\psi_{s_e,\m{X},s_m}]}=\eta_{[s_e,\m{X},s_m]}\sum_{\phi \in G[s_e,\m{X},s_m]}~_{\phi}\bra{s_m}~\otimes~_{\phi}\bra{s_e,\m{X}} 
\end{eqnarray}
We shall prove that for this basis state,
the only gauge invariant set of data is given by ($\overrightarrow{k},\overrightarrow{l},\overrightarrow{\m{X}}(v)$). To this end, it is sufficient to prove that:

(a) If $\ket{\gamma'_e,\overrightarrow{k'},\m{X'}}\otimes \ket{\gamma'_m,\overrightarrow{l'}}=\ket{\gamma_e,\overrightarrow{k},\m{X}}\otimes \ket{\gamma_m,\overrightarrow{l}}_{\phi}$, then $(\overrightarrow{k'},\overrightarrow{l'},\m{X'}(v'))=(\overrightarrow{k},\overrightarrow{l},\m{X}(v))$, where $v'=\phi(v)$;

(b) If $(\overrightarrow{k},\overrightarrow{l},\m{X}(v))=(\overrightarrow{k'},\overrightarrow{l'},\m{X'}(v'))$, where the two sets correspond to any two charge network states $\ket{\gamma_e,\overrightarrow{k},\m{X}}\otimes \ket{\gamma_m,\overrightarrow{l}}$ and $\ket{\gamma'_e,\overrightarrow{k'},\m{X'}}\otimes \ket{\gamma'_m,\overrightarrow{l'}}$, then there exists some gauge transformation $\tilde{\phi}$ such that $\ket{\gamma'_e,\overrightarrow{k'},\m{X'}}\otimes \ket{\gamma'_m,\overrightarrow{l'}}~=~\ket{\gamma_e,\overrightarrow{k},\m{X}}_{\tilde{\phi}}\otimes \ket{\gamma_m,\overrightarrow{l}}_{\tilde{\phi}}$.

Proof of (a):\\
$\ket{\gamma'_e,\overrightarrow{k'},\m{X'}}\otimes \ket{\gamma'_m,\overrightarrow{l'}}=\ket{\gamma_e,\overrightarrow{k},\m{X}}\otimes \ket{\gamma_m,\overrightarrow{l}}_{\phi}$ implies that:
\begin{eqnarray}\label{k-l}
k'_{x,\phi(\gamma_e)}&=& k_{\phi^{-1}(x),\gamma_e} \nonumber\\
l'_{x,\phi(\gamma_m)}&=& l_{\phi^{-1}(x),\gamma_m}
\end{eqnarray} 
For the background, we obtain:
\begin{eqnarray*}
\m{X}'(x)&=&\m{X}(\phi^{-1}(x))
\end{eqnarray*} 
When the point $x$ above coincides with a vertex $v$, we obtain:
\begin{eqnarray}\label{X}
&&\m{X}'(v)=\m{X}(\phi^{-1}(v))\nonumber\\
&\Rightarrow & \m{X}'(\phi(v))=\m{X}(v)\nonumber\\
&\Rightarrow & \m{X}'(v')=\m{X}(v)
\end{eqnarray}
where we have used the definition $v'=\phi(v)$. 
Now note that $(\overrightarrow{k},\overrightarrow{l})$ and $\overrightarrow{\m{X}}(v)$ have periodic and quasiperiodic extensions over the real line, respectively. Since any $\phi$ acts as a periodic diffeomorphism on the real line, and preserves the number of vertices of $\gamma_e$ and $\gamma_m$, any diffeomorphic image of the set $(\overrightarrow{k},\overrightarrow{l},\overrightarrow{\m{X}}(v))$ for some $\phi$ must be some cyclic permutation $(\overrightarrow{k},\overrightarrow{l},\overrightarrow{\m{X}}(v))$. Thus, from (\ref{k-l}) and (\ref{X}), we can conclude that the set $(\overrightarrow{k'},\overrightarrow{l'},\overrightarrow{\m{X'}}(v'))$ must be a cyclic permutation of $(\overrightarrow{k},\overrightarrow{l},\overrightarrow{\m{X}}(v))$, which in turn implies that the two sets are equal.

Proof of (b):\\
We choose a convention such that each edge $e_n$ ($n=1,...,N$) in a graph covers a semi-open interval $I_n=[x^{min}_n,x^{max}_n)$ such that $\cup_{n=1,..,N}~I_n=[0,2\pi)$. Thus, $\overrightarrow{k}=\overrightarrow{k_v}$ where $k_v$ denotes the value of $k$ at any vertex $v$. Since the number of components of $\overrightarrow{k}$ and $\overrightarrow{k'}$ are the same, the number of vertices in the two sets are also the same. A similar conclusion holds for the background embedding data $(\overrightarrow{\m{X}}(v),\overrightarrow{\m{X'}}(v'))$ and the matter data $(\overrightarrow{l},\overrightarrow{l'})$. Thus, it is always possible to find some $\phi$ such that: 
\begin{eqnarray}\label{k-phi}
&& k'_{v'}=k'_{\phi(v)}=k_{v},\nonumber\\
&& l'_{v'}=l'_{\phi(v)}=l_{v},\nonumber\\
&& \m{X}'(v')=\m{X}'(\phi(v))=\m{X}(v)
\end{eqnarray}

Now, let us define $\m{X}''$ as the image of $\m{X}$ under the action of some periodic diffeomorphism $\tilde{\phi}$. This implies:
\begin{eqnarray} \label{phi}
&&\m{X}''(x)=\m{X}(\tilde{\phi}^{-1}(x))\nonumber\\
&\Rightarrow & \m{X}''(v)=\m{X}(\tilde{\phi}^{-1}(v))
\end{eqnarray}
Since $\tilde{\phi}$ can be arbitrary, we are free to make the choice $\tilde{\phi}=\m{X}^{' -1} \circ \m{X}$, since $\m{X},\m{X}'$ are invertible. Now notice that for this choice, we have $\m{X}''(x)=\m{X}'(x)$. Using this in (\ref{phi}), we obtain: 
\begin{eqnarray}\label{phi-tilde} 
\m{X}'(v)=\m{X}(\tilde{\phi}^{-1}(v))
\end{eqnarray} 
Thus, using equations (\ref{phi-tilde}) and (\ref{k-phi}), we can conclude that it is always possible to find a $\tilde{\phi}$ such that the following holds:
\begin{eqnarray}
&& k'_{v'}=k'_{\tilde{\phi}(v)}=k_{v},\nonumber\\
&&l'_{v'}=l'_{\tilde{\phi}(v)}=l_{v},\nonumber\\
&& \m{X}'(\tilde{\phi}(x))=\m{X}(x) \Rightarrow \m{X}'(x)=\m{X}(\tilde{\phi}^{-1}(x))
\end{eqnarray}
This in turn implies that: 
\begin{eqnarray*}
\ket{\gamma'_e,\overrightarrow{k'},\m{X'}}\otimes \ket{\gamma'_m,\overrightarrow{l'}}~=~\ket{\gamma_e,\overrightarrow{k},\m{X}}_{\tilde{\phi}}\otimes \ket{\gamma_m,\overrightarrow{l}}_{\tilde{\phi}}
\end{eqnarray*}
This completes the proof that the complete set of gauge invariant data labelling a physical basis state is given by ($\overrightarrow{k},\overrightarrow{l},\overrightarrow{\m{X}}(v)$).
%


\begin{thebibliography}{99}
%
\bibitem{dirac} P. A. M. Dirac, Can. J. Math. {\bf 2}, 129 (1950); Lectures on Quantum Mechanics (Academic, New York, 1965).	
%
\bibitem{kuchar} K. V. Kuchar, Phys. Rev. {\bf D 39}, 1579 (1989). 
%
\bibitem{kuchar1} K. V. Kuchar, Phys. Rev. {\bf D 39}, 2263 (1989).
%
\bibitem{varadarajan} M. Varadarajan, Phys. Rev. {\bf D75}, 044018 (2007).
%
\bibitem{l} A. Laddha, Class. Quant. Grav. {\bf 24}, 4969 (2007).
%
%
\bibitem{laddha} A. Laddha and M. Varadarajan, Phys. Rev. {\bf D 78}, 044008 (2008).
%
\bibitem{laddha1} A. Laddha and M. Varadarajan, Classical Quantum Gravity {\bf 27}, 175010 (2010).
%
\bibitem{laddha2} A. Laddha and M. Varadarajan, Phys. Rev. {\bf D 83}, 025019 (2011).
%
\bibitem{ashtekar} A. Ashtekar, J. Lewandowski, Class. Quant. Grav. {\bf 14},  A55-A82 (1997);\\
A. Ashtekar, J. Lewandowski, Adv. Theor. Math. Phys. {\bf 1}, 388 (1998). 
%
\bibitem{rovelli} C. Rovelli, L. Smolin, Nucl. Phys. {\bf B} 442, 593 (1995); Erratum-ibid. {\bf B} 456, 753 (1995); \\
J. Lewandowski, Class. Quant. Grav. {\bf 14}, 71 (1997).
%
%
\bibitem{k} T. A. Koslowski, `Dynamical Quantum Geometry', e-Print: arXiv:0709.3465 [gr-qc].
%
\bibitem{s} H. Sahlmann, Class. Quant. Grav. {\bf 27}, 225007 (2010);\\
T. Koslowski, H. Sahlmann, SIGMA {\bf 8}, 026 (2012).
%
\bibitem{m} M. Varadarajan, e-Print: arXiv:1306.6126 [gr-qc].
%
\bibitem {almt} A. Ashtekar, J. Lewandowski, D. Marolf, J. Mourao, and T. Thiemann, J. Math. Phys. (N.Y.) {\bf 36}, 6456 (1995).
%
\bibitem{marolf} D. Marolf, in Proceedings of the 9th Marcel Grossmann Meeting, edited by V. G. Gurzadyan, R. T. Jantzen, and R. Ruffini (World Scientific, Singapore, 2002); available as D. Marolf, gr-qc/0011112.
%
%
\end{thebibliography}
\end{document}